\documentclass[12pt]{article}
\usepackage{amsmath}
\usepackage{amsfonts}
\usepackage{color}
\usepackage{graphicx}
\usepackage{epsf}
\usepackage{lscape}
\usepackage{amsfonts}
\usepackage{amsbsy}
\usepackage{amssymb}

\usepackage[T1]{fontenc}

\usepackage[greek,english]{babel}
\languageattribute{greek}{polutoniko}

\newcommand{\be}{\begin{equation}}
\newcommand{\ee}{\end{equation}}
\newcommand{\beqs}{\begin{eqnarray}}
\newcommand{\eeqs}{\end{eqnarray}}

\newcommand{\half}{{\frac{1}{2}}}

\newcommand{\sgn}{{\rm sgn}}

\def \nn {\nonumber}
\def \bb {\hskip -0.1cm}
\def \hb {\hskip -0.05cm}

\definecolor{light}{gray}{.80}
\definecolor{dark}{gray}{.40}
\definecolor{MidnightBlue}{cmyk}{1, 0.13, 0, 1}
\definecolor{sepia}{cmyk}{0, 0.83, 1 ,0.70}
\definecolor{sepial}{cmyk}{0, 0.43, .5 ,0.35}
\definecolor{red}{cmyk}{0, 1, 1 ,0}
\definecolor{yellow}{cmyk}{0, 0, 1 ,0}
\definecolor{goldenrod}{cmyk}{0,0.10, 0.84, 0}

\newcommand{\eq}[1]{{{{ \be #1 \nonumber\ee}}}}

\newbox\ncintdbox \newbox\ncinttbox
\setbox0=\hbox{$-$} \setbox2=\hbox{$\displaystyle\int$}
\setbox\ncintdbox=\hbox{\rlap{\hbox to
\wd2{\hskip-.125em\box2\relax \hfil}}\box0\kern.1em}
\setbox0=\hbox{$\vcenter{\hrule width 4pt}$}
\setbox2=\hbox{$\textstyle\int$}
\setbox\ncinttbox=\hbox{\rlap{\hbox to
\wd2{\hskip-.175em\box2\relax \hfil}}\box0\kern.1em}

\begin{document}

\renewcommand{\theequation}{\thesection.\arabic{equation}}
\csname @addtoreset\endcsname{equation}{section}
%
%
%
%
%
%
%
%
%
%
%
%
%
%
%
%
%
%
%
%
%
\begin{titlepage}
\null\vspace{-62pt} \pagestyle{empty}
\begin{center}
\rightline{June 2020}
\vspace{0.8truein} {\Large\bfseries
Exchange interactions, Yang-Baxter relations and transparent particles}\\
\vskip .15in
{\Large\bfseries ~}\\
{\bf {\large Alexios P. Polychronakos}}\\
\vskip .2in

{\itshape Physics Department, the City College of New York, NY 10031\\
and\\
The Graduate Center of CUNY, New York, NY 10016, USA}\\
\vskip .1in
{\fontsize{10pt}{10pt} apolychronakos@ccny.cuny.edu}

\fontfamily{cmr}\fontsize{11pt}{15pt}\selectfont
\vspace{.8in}
\centerline{\large\bf Abstract}
\end{center}
We introduce a class of particle models in one dimension involving exchange interactions that have
scattering properties satisfying the Yang-Baxter consistency condition. A subclass of these models
exhibits reflectionless scattering, in which particles are ``transparent'' to each other, generalizing a property
hitherto only known for the exchange Calogero model. The thermodynamics of these systems can be derived
using the asymptotic Bethe-ansatz method.

\end{titlepage}
\section{Introduction}

Integrable models in one space dimension have been an object of study in physics and mathematics for 
several decades. These include particle systems, such as the Lieb-Liniger model \cite{LiLi} and the Calogero class of models \cite{Calogero-1969,Sutherland-1971,Moser}
as well as spin systems such as the Heisenberg model \cite{Heisenberg} and the Haldane-Shastry model \cite{Haldane,Shastry}, and field theory systems such as the ``sine-Gordon" model and the principal chiral model.

Integrability is intimately related to the satisfaction of the Yang-Baxter (YB) equation \cite{Y,B}, which guarantees that 3-body
scattering can, in principle, be consistently derived from 2-body scattering. Although the YB equation is only a
necessary condition for integrability, in practice every system known to satisfy it turns out to be integrable. The discovery
of extended classes of systems that satisfy the YB equation is, thus, an interesting endeavor.

A special class of integrable models is many-body systems with particle-exchange interactions. This includes, primarily,
the exchange-Calogero model and its generalizations \cite{exop,Caloreview}. The advantage of such systems is that they provide a
compact proof of integrability and allow for the introduction of spins through a swap of particle and internal degrees of freedom. They also provide a convenient setting for operator-based techniques that apply eventually to the ordinary
Calogero model, such as, e.g., a convenient expression of the intertwining operator \cite{intert}.

A special and surprising property of the exchange-Calogero model is that, upon scattering, particles go through each
other with no backscattering \cite{SS,Caloreview}. In spite of the singular inverse-square interaction of the particles at coincidence points, the exchange part of their interactions ``conspires'' with their dynamical properties to
produce a complete and perfect ``teleportation'': the particles behave as if they were transparent to each other.
This greatly simplifies the
scattering properties of the system, trivially satisfying the Yang-Baxter equation and allowing for simple explicit
expressions for the energy eigenvalues in the asymptotic Bethe-Ansatz method.

The purpose of this note is to show that this property is not specific to the Calogero model but is shared by a very wide
class of models with exchange interactions.
Consequently, all these models are amenable to solution in the thermodynamic
limit via the asymptotic Bethe-Ansatz method. In fact, this class of models can be further extended such that is comprise models that {\it do} have backscattering but nevertheless satisfy the Yang-Baxter
equation and are, thus, equally amenable to solution. Whether all these models are integrable remains an open question,
with the answer being almost certainly in the negative. Nevertheless, their accessible and, to an extent, universal
properties in the thermodynamic limit, and their otherwise general interactions, render them attractive candidates for the
analytical description of various physical many-body systems.

\section{The exchange-Calogero model}

We briefly review the properties of the exchange-Calogero model to connect with known facts and set the stage for the more
general class of models to be introduced in the sequel.

Consider $N$ identical, but in principle distinguishable, nonrelativistic particles of unit mass in one dimension with coordinates $x_i$
and momenta $p_i$.  Define the ``exchange momentum'' operators $\pi_i$ as
\be
\pi_i = p_i + \sum_{j (\neq i)} \frac{i \ell}{x_{ij}} M_{ij}
\ee
where $\ell$ is a real constant and $M_{ij}$ are position-exchange operators satisfying
\be
M_{ij} A_i = A_j M_{ij}~,~
M_{ij} A_k = A_k M_{ij} ~~ (k\neq i,j)
\ee
with $A_i$ any one-particle operator depending on the $\{x_j\}$, $\{p_j\}$ and $\{ M_{ij} \}$.

The above $\pi_i$ are Hermitian one-particle operators satisfying
the important and nontrivial relation
\be
[ \pi_i , \pi_j ] = 0
\ee
Choosing a `free' Hamiltonian in the $\pi_i$
\be
H = \sum_{i=1}^N \half \pi_i^2 = 
\sum_i \half p_i^2 + \sum_{i<j} \frac{\ell (\ell -M_{ij} )}
{x_{ij}^2}
\ee
we obtain a Calogero-like model with exchange interactions.

It is obvious that all operators involving only $\pi_i$ will commute.
In particular, the permutation symmetric quantities
\be
I_n = \sum_{i=1}^N \pi_i^n ~,~~~ [ I_n , I_m ] = 0
\ee
which include the Hamiltonian as $I_2 = 2H$, constitute $N$ independent conserved quantities in involution,
proving the integrability of the exchange-Calogero system.
We can also consider the bosonic or fermionic sectors of this model, on which $M_{ij} \Psi = \pm \Psi$. On these sectors the Hamiltonian becomes the one of the standard Calogero model
\be
H =
\sum_i \half p_i^2 + \sum_{i<j} \frac{\ell (\ell \mp 1)} {x_{ij}^2}
\ee
while the correspondingly projected $I_{n\pm}$ become the standard Calogero integrals of motion.

Similar
exchange-augmented models can be defined for the harmonic (confining) Calogero, the periodic (Sutherland) and the
hyperbolic models, with similar but somewhat extended algebraic properties.
We can also introduce internal degrees of freedom (spins)
by considering other permutation sectors of the exchange model, which will be relevant later on.

The exchange-Calogero model has the surprising property that scattering particles go through each other without
backscattering (reflection). This is almost immediate in the exchange-operator formulation: consider a simultaneous eigenstate
of the commuting exchange-momenta $\pi_i$ with eigenvalues $k_i$. At asymptotic regions of the configuration space where 
$|x_i - x_j | \to \infty$ the interaction terms in $\pi_i$ drop and $\pi_i \to p_i$. Therefore, the wavefunction is an asymptotic
scattering state with asymptotic momenta $k_i$. Note that this is true in {\it  all} asymptotic regions for {\it any} ordering of the
$x_i$. Therefore, there is no momentum mixing before and after scattering. Particles emerge with the same momentum
after their scattering as the one before and there is no reflected component. The scattering has complete transmission, with at
most an overall phase shift. The sector in which the coordinates $x_i$ are in the inverse ordering as the corresponding
scattering momenta $k_i$ is the incoming wavefunction while the one with the same ordering is the outgoing wavefunction,
other sectors representing intermediate stages of scattering.

It is instructive to consider the two-body scattering and rederive the above result using the details of the dynamics. For the
ordinary (non-exchange) Calogero model the potential is impenetrable and there is complete reflection with a phase shift
equal to $\pi \ell$, where $\ell(\ell-1)$ is the strength of the interaction and $\ell >0$. So for the bosonic sector of the model
with scattering momenta $k_1 > k_2$ the phase shift is $\pi \ell$, while for the fermionic sector, with strength $\ell(\ell+1) =
(\ell+1)[(\ell+1) -1]$, it is $\pi (\ell+1)$.

A scattering of two particles with incident momenta $k_1 > k_2$ at $x_1 < x_2$ will have an outgoing (scattered)
wavefunction at $x_1 > x_2$ with momenta $k_1 , k_2$ as well as a reflected wavefunction at $x_1< x_2$ with momenta
$k_2 , k_1$. We can express this state as the sum of a fermionic state and a bosonic state such that the $k_2 , k_1$
part of the wavefunction vanish for $x_1 > x_2$. The crucial fact is that at $x_1 < x_2$ the fermionic and bosonic wavefunctions
with $k_2 , k_1$ interfere with the {\it same} phase as for $x_1 > x_2$: they have one extra relative phase $\pi$ because of the
difference in scattering phases and an additional phase $\pi$ because of the opposite
sign in their exchange properties. So the $k_2 , k_1$ wavefunction will vanish for $x_1 < x_2$ as well: there is {\it no}
backscattered (reflected) wave.

The key lesson from the above two-body scattering discussion is that, if the fermionic and bosonic reflection phase shifts
differ by $\pi$, then the scattering is reflectionless.

\section{Scattering phases and Yang-Baxter relations}

Before introducing the generalized models, we review the basic facts about the scattering matrix, the transfer matrix
and the Yang-Baxter relation.

Consider a two-particle scattering process. Assuming translation invariance and factoring out the center of mass,
we end up with a one-dimensional scattering in the relative coordinate $x$. For Hamiltonians that become free in the
asymptotic regions $x \to \pm \infty$ the asymptotic form of the wavefunction will be (with $k>0$)
\beqs
\psi (x) &= A_{L}\, e^{i k x} + B_{L}\, e^{-i k x} ~,~~~ &x<0 \nonumber \\
&= A_{R} \, e^{i k x} + B_{R}\, e^{-i k x} ~,~~~ &x>0 
\eeqs
The relation between the four constants $A_{_{L,R}} , B_{_{R,R}}$ can be expressed either in terms of the transfer
matrix $U$, relating left and right amplitudes
\be
\psi_{R} = U \psi_{L} ~~ \Rightarrow ~~
\left[\begin{matrix} A_{R} \cr B_{R} \end{matrix} \right] = \left[\begin{matrix} {U_{AA} ~~ U_{AB}} \cr {U_{BA} ~~ U_{BB}} \end{matrix} \right] \left[\begin{matrix} A_{L} \cr B_{L} \end{matrix} \right] 
\ee
or in terms of the scattering matrix $S$, relating incoming and outgoing amplitudes
\be
\psi_\text{in} = S \psi_\text{out} ~~ \Rightarrow ~~
\left[\begin{matrix} A_{R} \cr B_{L} \end{matrix} \right] = \left[\begin{matrix} {T_R ~~ R_R} \cr {R_L ~~ T_L} \end{matrix} \right] \left[\begin{matrix} A_{L} \cr B_{R} \end{matrix} \right] 
\ee
$T_{L,R}$ and $R_{L,R}$ are the left and right transmission and reflection amplitudes. Unitarity $S^\dagger = S^{-1}$
imposes the conditions
\beqs
|T_R |^2 = |T_L |^2 &,& ~~|R_L |^2 = |R_R |^2 \nn \\
|T_L|^2 + |R_L|^2 = 1 &,&~~ T_R^* R_R + R_L^* T_L = 0
\eeqs
In addition, permutation invariance $H M_{12} = M_{12} H$, which translates to parity invariance for the relative Hamiltonian,
implies
\be
T_L = T_R \equiv T ~,~~~ R_L = R_R \equiv R
\ee
with $T$ and $R$ satisfying
\be
|T|^2 + |R|^2 = 1 ~,~~ T R^* + R T^* = 0
\ee
Correspondingly, the transfer matrix elements are expressed as
\be
U_{BB} = U_{AA}^* = \frac{1}{T} \, ,~~ U_{AB} = U_{BA}^* = \frac{R}{T}
\ee
The elements of the transfer or scattering matrix can be expressed in terms of two phases, the bosonic and fermionic scattering
phases $\theta_+$ and $\theta_-$. A bosonic (fermionic) state satisfies $A_R = \pm B_L$, $B_R = \pm A_L$ with $+(-)$
respectively, leading to
\be
A_R =e^{i\theta_\pm} A_L ~,~~ T \pm R = e^{i\theta_\pm}
\ee
and we define $\theta_\pm (-k) = - \theta_\pm (k)$ such that the above relation hold also for $B_{R}$ and $B_{L}$.
(Note that the above are the {\it transmission} phase shifts. In the previous section we used the {\it reflection} phase shifts, relating
$A_L$ and $B_L$, which are $\theta_+$ and $\pi + \theta_-$.)
Consequently
\be
T(k) = {e^{i\theta_+ (k)} \hb +\hb e^{i\theta_- (k)} \over 2} = T(-k)^* ,~ R(k) = {e^{i\theta_+ (k)} \hb -\hb e^{i\theta_- (k)} \over 2} = R(-k)^*
\ee
and the transfer and scattering matrices are expressed in terms of the asymptotic momentum exchange operator 
(relative asymptotic momentum inversion operator) $P_{12}$ and the coordinate exchange operator$M_{12}$
\be
U = {1 \over T} + {R \over T} P_{12} =  \frac{2}{e^{i\theta_+ (k)} + e^{i\theta_- (k)}} + \frac{e^{i\theta_+ (k)} + e^{i\theta_- (k)}}{e^{i\theta_+ (k)} + e^{i\theta_- (k)}} P_{12}
\label{sca.1}\ee
\be
S = T + R M_{12} = e^{i \theta_+ (k)} {1+M_{12} \over 2} + e^{i\theta_- (k)} {1-M_{12} \over 2}~~~~~~
\label{sca.2}\ee

Now consider a three particle scattering. The asymptotic region where all particles are far apart consists of
6 sectors according to the ordering of the particle coordinates, namely $(123) = \{x_1\,$$<\,$$x_2\,$$<\,$$x_3 \}$ and permutations,
separated by the coincidence planes $x_1 = x_2$, $x_1 = x_3$ and $x_2 = x_3$.
Assume that an energy eigenstate exists such that the wavefunction in one of the asymptotic regions is a combination of
plane waves with momenta $k_1 , k_2 , k_3$ and their permutations. Then the state will retain this form in all sectors, the
wavefunctions in adjacent sectors related through the relevant transfer matrix, denoted $U_{ij}$ when it relates particles
$i$ and $j$ leaving the third one alone. Following the wavefunction from sector $(123)$ to $(321)$ 
\begin{figure}
\vskip -3cm\hskip -1.1cm\includegraphics[scale=0.5]{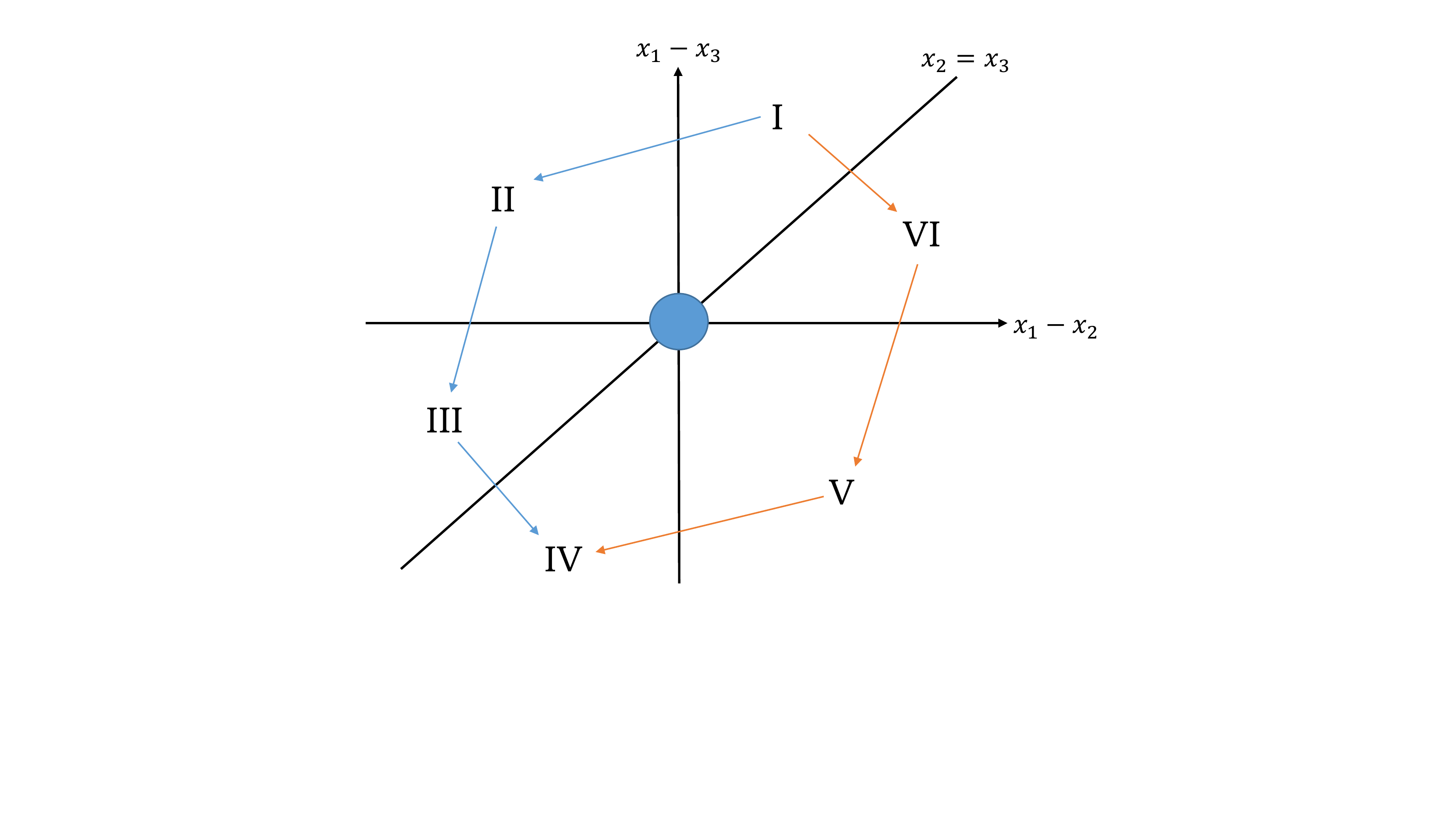}
\vskip -2cm
\caption{\small{Scattering in the space of relative coordinates of three particles. Transferring from region I ($x_1\bb
>\hb x_2 \bb >\hb x_3$) to region IV ($x_3\bb >\hb x_2\bb >\hb x_1$) via path {\color{blue}{$I \to II \to III \to IV$}} should
give the same wavefunction as via path {\color{magenta}$I \to VI \to V \to IV$}}. The Yang-Baxter condition can be interpreted as the absence
of Aharonov-Bohm effects from the triple coincidence line $x_1 = x_2 = x_3$.}
\end{figure}
\beqs
&(123) \to (213) \to (231) \to (321) ~\Rightarrow~ &\Psi(321) = U_{23}\, U_{13} \,U_{12} \Psi(123) \nn \cr
&(123) \to (132) \to (312) \to (321) ~\Rightarrow~ &\Psi(321) = U_{13} \,U_{13}\, U_{23} \Psi(123)
\eeqs
If we want such an energy eigenstate to exist for {\it arbitrary} combinations of plane waves in sector $(123)$ then we must have
\be
U_{23}\, U_{13} \,U_{12}  = U_{13} \, U_{13}\, U_{23}
\label{YB}\ee
This is the Yang-Baxter (YB) relation expressed it terms of the transfer matrix rather than the more familiar (and equivalent) expression
in terms of the scattering matrix. It expresses the absence of ``Aharonov-Bohm'' effects in scattering around the triple
coincidence line $x_1 = x_2 = x_3$ which renders the configuration space non-simply connected (Fig. 1).

The YB relation is a {necessary} condition if a {\it generic} superposition of plane waves in the asymptotic region is to be
an energy eigenstate. States that satisfy the relation $U_{23}\, U_{13} \,U_{12} \Psi = U_{13} \, U_{13}\, U_{23} \Psi$ even if
(\ref{YB}) is not satisfied are still viable candidates. Generically, such states are only bosonic or fermionic wavefunctions, for which
$U_{ij}$ and $S_{ij}$ become essentially scalar operators (in particular, $S = e^{i\theta_\pm}$). Such states, however, do not span the
full set of solutions for three or more distinguishable particles.

Integrable models have the defining property that three-body scattering can be expressed in terms of two-body scattering, and thus
asymptotic momenta do not spread and are conserved quantities. For such models the YB equation is a {\it necessary} condition.
However, it is not a {\it sufficient} condition for integrability. In general the scattering
will contain three- and higher-body scattering corrections, arising from the regions of the configuration space in the vicinity of triple or higher
coincidence planes. It is, nevertheless, an interesting fact that all models known to satisfy the YB relations are also integrable.

 \section{Generalized models} 

In analogy to the exchange-Calogero model, define the operators
\be
\pi_i = p_i + i\sum_{j (\neq i)} W(x_i - x_j ) M_{ij} 
\ee
with a general `prepotential' function $W(x)$ and take as Hamiltonian
\be
H = \sum_{i=1}^N \half \pi_i^2
\ee
Hermiticity of the $\pi_i$ requires $W(x)^* = -W(-x)$, while elimination of terms linear in momenta in the Hamiltonian
requires $W(x) = - W(-x)$, so overall $W(x)$ must be real and antisymmetric
\be
W(x) = -W(-x) = W(x)^*
\ee
Under these conditions the Hamiltonian and commutation relations of the $\pi_i$ become
\be
[ \pi_i , \pi_j ] = \sum_{k \neq i,j} (W_{ij} W_{jk} + W_{jk} W_{ki} + W_{ki} W_{ij} ) M_{ij} (M_{jk} - M_{ik} )
\label{commpi}\ee
\be
H = \sum_i \half p_i^2 + \sum_{i<j} (W_{ij}^2 + W'_{ij} M_{ij} ) \,+\bb \sum_{i,j,k\, \text{\tiny distinct}} \bb\bb W_{ij} W_{jk}\, M_{ij} M_{ik}
\ee
where we adopted the shorthand $W_{ij} = W( x_i - x_j )$. In the case of the Calogero model the three-body terms above vanish or
become a constant. For general $W(x)$ they survive, but are irrelevant for two-body scattering. In fact, we will simply omit
them for the considerations in this analysis.

To support scattering, $W(x)$ has to become constant at infinity:
\be
W(x \to \pm \infty) = \pm a
\ee
Consider a two-particle scattering (all the other particles being far away).
Separating center-of-mass and relative coordinates, the Hamiltonian for the relative motion is
\be
H_r = \bigl[ p + i W(x) M \bigr]^2 ,~~ x = x_1 \bb -\hb x_2 \,,~p = \frac{p_1 \bb -\hb p_2}{2} , ~ M=M_{12}
\ee
The scattering equation for bosonic or fermionic wavefunctions with asymptotic momentum $k$ reads
\eq{
\left[ p \mp i W(x) \right] \left[ p \pm i W(x) \right] \psi_\pm (x) = (k^2 + a^2 ) \psi_\pm (x)
}
From the above ``supersymmetric'' form, acting with $p \pm i W(x)$ on both sides, we deduce that the two
scattering wavefunctions are related as
\be
\psi_- (x) \sim \left[ p + i W(x) \right] \psi_+ (x)~,~~\psi_+ (x) \sim \left[ p - i W(x) \right] \psi_- (x)
\ee
and from the asymptotics at $x \to \infty$ we deduce
\be
e^{i \theta_+ (k)} = \frac{k-ia}{k+ia} \, e^{i\theta_- (k)}
\ee
From this and (\ref{sca.1},\ref{sca.2}) we finally obtain our main result
\be
T(k) = \frac{k}{k+ia}\, e^{i\theta_- (k)} ~,~~~ R(k) = -\frac{ia}{k+ia}\, e^{i\theta_- (k)}
\label{tran}\ee

We observe that for the class of models with prepotential vanishing at infinity ($a=0$) we have {{no reflection}}: particles are ``transparent''
to each other in scattering. They simply pick up a scattering phase $\theta_+ (k) = \theta_- (k)$.

For the more general case $a\bb \neq\bb 0$, $R(k)$ does not vanish and there is backscattering. Nevertheless, the scattering matrix
\be
S_{ij} = e^{i \theta_- (k_{ij} )} \left( {k_{ij} \over k_{ij} + i a} - {i a \over k_{ij} + i a} M_{ij} \right)
\label{scata}\ee
 {\it{still}} satisfies the Yang-Baxter equation. Asymptotic momenta can spread (if at all) only in three-body scattering.

It is instructive to consider several examples.

a. $W(x) = a \, \rm{sgn}(x)$ produces a two-body potential $V_{ij} = a^2 + 2a \delta(x_{ij}) M_{ij}$. Since the $\delta$-function term
only matters when $x_i = x_j$ the permutation operator acts trivially and thus, up to an uninteresting constant $a^2$, the potential
is $V_{ij} = 2a \delta (x_i - x_j )$. This is the Lieb-Liniger model with pairwise contact interactions. For fermionic states vanishing
at coincidence points the potential is irrelevant and we trivially have $\theta_- = 0$. So (\ref{scata}) reproduces the well-known
scattering matrix of the model. Since triple-scattering points are of measure zero for contact interactions, the two-body scattering
reproduces the full dynamics of the model and we have a complete proof of integrability. 

b. $W(x) = \ell / x$ reproduces the exchange-Calogero model where transparency
can be shown directly from the commutativity of the exchange-momenta $\pi_i$. It is also obvious from (\ref{tran}) and
the fact $W(\pm \infty) = 0$.

c. $W(x) = \ell \, \coth x$ gives the hyperbolic exchange-Calogero model where there is indeed backscattering and a nontrivial scattering
phase shift.

Other models can be ``cooked up'' by suitable choices of $W(x)$. E.g.,

d. $W(x) = \ell (x + \sgn(x) \, a)^{-1}$ gives $V_\pm (x) = \ell(\ell \mp 1)/
(|x| + a)^2 - (2\ell/a) \delta(x)$, a `volcano-like' potential leading to reflectionless scattering.

e. $W(x) = \ell /x + \sgn(x) \, b$ gives $V_\pm (x) = \ell(\ell\mp 1)/x^2 +2b\ell/|x| +2\ell b \delta (x)$. The $\delta$-function potential
is actually irrelevant, since the wavefunction will vanish at $x=0$ (even in the bosonic case), so this is a Calogero plus Coulomb interaction
with a nontrivial scattering and phase shift.

\section{Thermodynamic Bethe-ansatz solution}\label{ABA}

All the above models, though not necessarily integrable, are amenable to an asymptotic Bethe ansatz
treatment of their thermodynamics. The most straightforward case is the reflectionless one ($a=0$) where there
is a single scattering phase $\theta (k) = \theta_+ (k) = \theta_- (k)$.

The asymptotic Bethe-ansatz (ABA) form of the wavefunction with asymptotic momenta $k_1 > \dots >k_N$ is
\be
\Psi = e^{{i\over 2} \sum_{i<j} \vartheta (x_i - x_j) \theta (k_i - k_j )} e^{i\sum_i k_i x_i}
\label{kasi}\ee
where $\vartheta (x) = (1+\sgn (x))/2$ is the step function. It is in the form of a single plane wave with appropriate
phase shifts in each asymptotic permutation sector. It is an energy eigenstate with energy
\be
E = \sum_i \half k_i^2 
\label{E}\ee
In principle there are $N!$ different degenerate asymptotic eigenstates differing by a permutation of the $x_i$
appearing in (\ref{kasi}).

In a periodic space of length $L$, assuming the above asymptotic form of the wavefunction is a valid approximation,
periodicity in all $x_i$ imposes the conditions on the Bethe-ansatz momenta
\be
L k_i = 2\pi n_i + \sum_{j(\neq i)} \theta(k_i - k_j )
\label{Bet}\ee
\noindent
The $n_i$ are integer quantum numbers, ordered as $n_1 \ge \dots \ge n_N$ in order to preserve the ordering of
$k_i$. At face value, this eigenstate would be $N!$-degenerate due to the $N!$-fold degeneracy of (\ref{kasi}).
However, this is not true when two or more of the $n_i$ coincide: continuity from the non-interacting case (by, say,
choosing as prepotential $W_\lambda = \lambda W(x)$ and taking $\lambda \to 0$) suggests that states differing by
permuting the particles at coincident quantum numbers $n_i$ are the same. (The distinct asymptotic forms (\ref{kasi})
in this case are approximations of the unique exact energy eigenstate.) The end result is that the degeneracies are the
same as those of $N$ distinguishable particles occupying the discrete states $n_i$. 

We can derive the thermodynamic grand potential $\Omega$ of the system, assuming that it consists of identical particles coming in species $a=1,\dots, M$, each one of quantum statistics $s_a$, with $s_a = 1\, (-1)$ for bosons (fermions). For an inverse
temperature $\beta$ and chemical potentials $\mu_a$
\be
\beta \Omega = \sum_a \int \frac{Ldk}{2\pi}\, {s_a}
\ln \left[ 1 - s_a \,e^{-\beta (\epsilon (k)-\mu_a )}\right]
\ee
The effective single-state energy $\epsilon (k)$ is a continuum version of (\ref{E}) and the Bethe equations (\ref{Bet})
taking into account the other particles in the systems, and is determined from
\be
\epsilon (k) = \half k^2 + \beta^{-1} \sum_a \int \frac{dq}{2\pi} \theta' (k-q) \, {s_a}
\ln \left[ 1 - s_a \, e^{-\beta (\epsilon (k)- \mu_a )}\right]
\ee
The approximate thermodynamic properties of our generalized systems are derived from the above equations. In particular,
the second virial coefficient, involving only two-particle interactions, is expected to be reproduced exactly.

\section{Conclusions}

The refrectionless nature, or Yang-Baxter consistency, of the generalized family of models exposed here is suggestive of
other special properties of these systems. The obvious question is whether these models are integrable. This is rather
unlikely, as it would imply the existence of a vast manifold of integrable systems. The question is even not well posed at this
level, as the nature of three- and higher-body potentials, which would be crucial for integrability, has not been specified.
The most natural choice for such terms
would be the one implied by the form $\half \sum_i \pi_i^2$ for the Hamiltonian, but a direct proof (or disproof) of
integrability based on the generalized commutation relations (\ref{commpi}) of the $\pi_i$ is lacking.

The possibility to introduce spins in the above models is essentially immediate, trading the distinguishable
nature of particles for corresponding internal degrees of freedom. In the standard construction, choosing overall
bosonic or fermionic states swaps the coordinate exchange operators $M_{ij}$ for corresponding spin exchange operators
$\sigma_{ij}$, endowing the particles with spins and ferromagnetic or antiferromagnetic spin interactions.
The ABA energy eigenstates of this model are deduced form the general construction of section {\bf \ref{ABA}}, with the
spin content as implied by putting the spins of particles at the same $n_i$ in totally symmetric (for bosons) or
antisymmetric (for fermions) representations. Confining the particles on a ring and driving the interaction potential to infinity
(by using the prepotential $W_\lambda = \lambda W(x)$ and taking $\lambda \to \infty$) freezes the particle positions on a
lattice and decouples the spin and kinematical (vibrational) degrees of freedom. A similar construction based on the present
generalized exchange models would lead to new interacting spin chain models, solvable if the spectrum of the full kinematical
models can be reliably calculated by the ABA ansatz.

The most interesting physical application of the above models would lie in their (approximate) description of realistic physical
systems and
the corresponding access to the thermodynamics of these systems they would afford. This remains an open subject for
investigation.
\vskip 0.6cm

\noindent
{\bf{\large Acknowledgment}}
\vskip 0.2cm
\noindent
This research was partially supported by NSF under grant 1519449.

\end{document}